# Public speech recognition transcripts as a configuring parameter in human-agents interactions

Damien Rudaz and Christian Licoppe

*Abstract*— Displaying a written transcript of what a human said (i.e. producing an "automatic speech recognition transcript") is a common feature for smartphone vocal assistants: the utterance produced by a human speaker (e.g. a question) is displayed on the screen while it is being verbally responded to by the vocal assistant. Although very rarely, this feature also exists on some "social" robots which transcribe human interactants' speech on a screen or a tablet. We argue that this informational configuration is pragmatically consequential on the interaction, both for human participants and for the embodied conversational agent. Based on a corpus of co-present interactions with a humanoid robot, we attempt to show that this transcript is a contextual feature which can heavily impact the actions ascribed by humans to the robot: that is, the way in which humans respond to the robot's behavior as constituting a specific type of action (rather than another) and as constituting an adequate response to their own previous turn.

## I. INTRODUCTION

A basic property of daily human-human interactions is that what is "inside people's head" is not a publicly available resource for participants involved in a local situation [1], [2]. As humans "don't carry MRI machines with them out in the world" [3], how a participant hears, understands, judges, etc. its interlocutors' actions is not a directly accessible information for these interlocutors: there exists no "internalist" window allowing humans involved in a conversation to immediately witness other participants' cognitive processes in real time. Most of the time, interactants can only rely on other participants' responses (verbal and embodied, including facial expressions) to identify if and how their own prior action was understood (as a question about X, as a request, as a greeting, etc.) [4], or if their words or gestures were even perceived by their interlocutor. Whether for the involved actor or the researcher studying video data, one must always rely on "inferential procedures" [5] to establish relationships between "discourse and cognition" [5].

In particular, as a matter of fact, human recipients do not display on their foreheads the exact words they hear (and potentially mishear) during other humans' speaking turns. A somewhat obvious consequence of this state of affairs is that, when, from an internalist point of view, an interlocutor completely mishears another participant's speaking turn, this is not an accountable phenomenon of miscommunication in itself: the only available resource for a co-present participant to detect and repair a potential miscommunication is this interlocutor's embodied or verbal response to the previous turn [6]. Yet, this specific informational ecology of human-human interactions appears not to translate entirely to a substantial part of human-agent interactions: those where the agent provides a written trace of what it heard the human say, i.e. a transcript of the speech recognition.

For example, on the commercial humanoid robot Pepper, the nominal behavior is to display on its tablet (attached to its torso) a transcript of what human interlocutors are saying, as heard by the robot. The top of this robots' belly screen features a "speechbar" where the result of the automatic speech recognition will be written, once the robot hears no more speech during more than 200ms (cf. fig.1). Hence, a Pepper robot which was just greeted with "hello" will display "hello" at the top of its tablet before it starts its return greeting action. Pepper's official documentation explicitly states that the speech recognition transcript (as part of the "speechbar") is intended as a resource towards which participants can orient to get a better grasp of the situation when interacting with Pepper, i.e. "to understand if the robot is listening, hearing something and what has been understood"[1]. In this sense, the transcript is expected to clarify the robot's own understanding (or misunderstanding) of the situation. Similar transcriptions of the "audio speech recognition" can be found on smartphone vocal assistants (e.g. Siri, Google Assistant, Bixby, etc.), where the vocal response of the assistant will be produced after the display of what a user said. In this view, human participants have direct access to the exact receipt of the words they have pronounced before any other return action from the agent can be achieved.

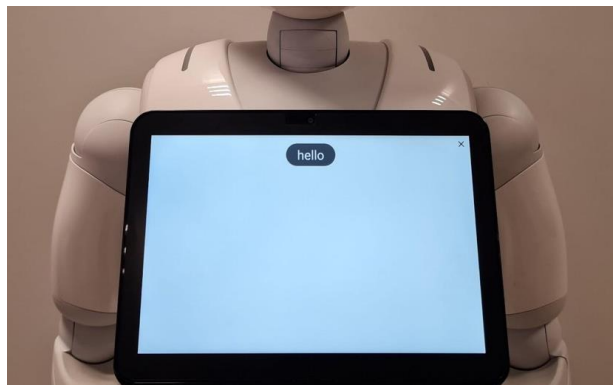

Figure 1. The Pepper robot transcribes "hello" after a human uttered this word

We will present two typical ways in which the tablet was used as a resource by participants *when the robot "misheard" the previous turn of a human*. These troublesome exchanges

D. Rudaz is with Telecom Paris and the Institut Polytechnique de Paris, Palaiseau, France (e-mail: damien.rudaz@telecom-paris.fr).

C. Licoppe is with the Telecom Paris and the Institut Polytechnique de Paris, Palaiseau, France (e-mail: christian.licoppe@telecom-paris.fr).

[1] Official documentation is available here:
https://qisdk.softbankrobotics.com/sdk/doc/pepper-sdk/ch4_api/conversation/conversation_feedbacks.html#conversation-feedbacks

appear more likely to reveal participants' orientation towards the Automatic Speech Recognition Transcript (thereafter ASR) by contrast with perfectly smooth interactions. We argue that the resulting informational ecology is consequential on the way miscommunications are detected and dealt with – compared with other commercial robots not relying on a transcription, or even with human-human interactions. In particular, we attempt to show specific situations in which the presence of a speech recognition transcript limited the sequential plasticity [7] of the robot's actions and, more generally, the conversational "moves" which were available to it.

## II. METHOD

To investigate the impact of this feature in situated interactions, we base our analysis on a video corpus recorded in July 2022 in Paris, at the Cité des Sciences et de l'Industrie, one of the biggest science museums in Europe. The first half of this corpus consists of 100 naturally occurring interactions with a Pepper robot placed in the hall of this museum. The second half is composed of 100 additional interactions which occurred with the same robot in a laboratory open to the public at the Cité des Sciences et de l'Industrie, where participants were asked to wear an eye tracking device. In both cases, participants (groups or single individuals) interacted with the same Pepper robot, on which was running a chatbot designed to converse on a wide variety of topics. The tablet attached to Pepper's torso was entirely blank, except for the speech recognition transcripts which appeared at the top of it. Recorded interactions were analyzed qualitatively, following an Ethnomethodology and Conversational Analytic methodology (thereafter EMCA).

## III. FRAGMENTS

Embodied actions were transcribed following [8] 's multimodal transcription conventions[2]. Transcription conventions can be found in appendix. P1 and P2 respectively refers to Participant 1 and Participant 2. RO refers to the robot. Translated from French.

### A. Fragment 1

```
1.        @(0.3)
   p1     @leans towards RO-->
2. P1  what is your job?
3.        (0.5)@(1.1)%(0.2)
   p1          -->@
   ro              %displays "what is your mel"-->
4. RO  I heard£#       what is your           £
   p1         £throws head backward and right£
   fi         #fig.2
```

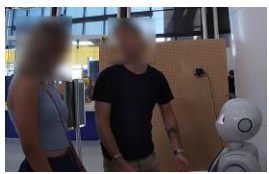

Figure 2.  P1 turns away from RO after reading the ASR transcript.

```
5. RO  m[el ^ (.) but] I don't know what to respond
   p1      ^walks away from RO-->
6. P1    [(    )    mel]
7.        (0.3)%(0.7)@(0.9)
   ro        -->%
   p2              -->@leans towards RO-->
```

---
[2] https://www.lorenzamondada.net/multimodal-transcription

```
8. P2  what: is (.) your (.) job:?
9.        (0.3)%(0.3)
   ro        %displays "what is your friend"->
10.RO  I've got@ two s[up*er]# @ fri+ends% (.)
   p2          -->@-straightens up-@
   p2                  *points tablet-->
   p2                          +gazes P2
   fi                          #fig.3
```

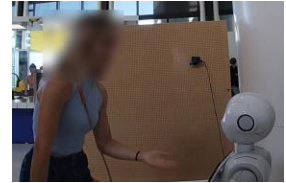

Figure 3.  P2 points at the robot's tablet

```
11.              [thh]
12.RO  ^(.) nao and romeo (.)
   p2  ^steps towards P1-->
13.RO  +(.)[all^ humans]% and robot are ^my friends too
   p2  ->+gazes at RO>>
   p2       -->^----steps towards RO----^
   ro                   -->%
14.P2     [((laughs))]
```

This fragment contains two similar treatments of the robot's conduct as inadequate, both of which are initiated before the robot's vocal and gestural response. While RO's answer (L.4) is still underway, P1 throws his head backwards and walks away from RO (L.5), as an embodied negative evaluation of the robot's conduct: when P1 initiates this embodied reaction (L.4), RO has so far only uttered "I heard". P1 then publicly singles out the troublesome word ("mel") by repeating it out loud (L.6), in overlap with RO's verbal answer. The robot's gestural and verbal response is therefore not responded to nor looked at by P1. P2 attempts to repair this interaction by formulating the same inquiry but, this time, with longer pauses between each word (L.6). Unlike what it did after P1's inquiry, RO responds to P2 without verbally indicating what it heard (L.10), but still displays an ASR transcript (L.9). As P1 did just before, P2 starts to react to RO's conduct as troublesome and noticeable when no meaningful content has yet been made hearable in RO's ongoing verbal answer: RO has only uttered "I've got" (L.10) when P2 starts to straighten up. She then points towards the belly screen, laughs, and produces a participation shift by gazing and stepping towards P1 while speaking with him. Significantly, while P2 is reacting to the ASR transcript, she disregards the vocal response that RO keeps producing: during RO's turn, she shifts her attention towards P1 and speaks or laughs in overlap with RO's hearable response. This supports an interpretation where the transcribed information has priority over the verbal response of the robot as an "answerable" or, at least, as a "noticeable" conduct.

### B. Fragment 2

```
1. P1  can you look up in:: the air?
2.       (2.9)%(0.3)
   ro        %displays "can you clean up the head upward"-->
3. RO  sorry (.) I do not have this feature.
4.       (0.3)%(1.9)
   ro    -->%
5. P1  °well I don't know°@ -it scares me a little bit* hh
   p1                    @turns towards P2-->
6.       (1.6)
```

```
7. P2 you're @not speaking loud enough () by clean up (it
meant) look up
   p1      -->@turns towards RO-->
8.    (2.0)
9. P1 it @wrote clean up?#
   p1 -->@turns towards P2-->
   fi                   #fig.4
```

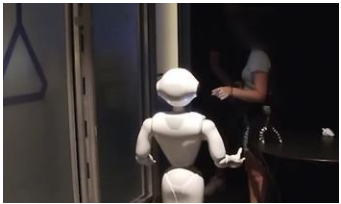

Figure 4. P1 points the robot's tablet after asking P2 for confirmation

```
10.   (1.0)£(0.5)£@(0.5)
   p2      £nods£
   p1            @turns and leans towards RO>>
11. P1 can you >look< up?
```

Two alternative understandings of the same scene coexist in this fragment. Each of them rely on a different reading of the speech recognition transcript (indicated in green) by P1 and P2. At first, even though RO's utterance denies ("I do not have this feature", L.3) what P1 was asking for (L.1), P1 does not make this request relevant as an adequate second pair part to her own turn: she does not attempt to repair the interaction, nor characterizes RO's response as a breach of the conditional relevance established by her question. In other words, RO's answer is *sequentially relevant* (as a response to a request) even though it halts the progress of the activity at hand (by rejecting the request). From the external point of view of the analyst, who has access to what the robot heard, the beginning of this fragment therefore displays an example of a misunderstood speaker who "do not grasp that they have been misunderstood" [9]. Yet, conversely, P2 characterizes RO's answer as responding to a turn which was not produced by P1: by informing P1 that RO transcribed "clean" instead of "look up" (L.7), and by stating that she didn't "speak loud enough" (L.7), he treats P2's original question as a source of trouble connected to RO's mismatched transcript.

Significantly, once information about what was written on the ASR transcript is provided to P1 by P2, this completely reconfigures, a posteriori, the action attributed to RO by P1. By asking a confirmation question ("it wrote clean?", L.9), P1 manifests she is dealing with new contextual information, on which she didn't previously rely to interpret RO's conduct. This is confirmed when, after this turn, she immediately casts the robot's previous answer as inadequate by repeating her own original question (L.11) – while insisting on the term which was stated by P2 to be mistranscribed. This reevaluation illustrates how much the social meaning of verbal and embodied responses from the robot can depend upon what the robot transcribes on its tablet: here, the ASR transcript is used as a resource to reevaluate the relevance of the second pair part produced by the robot.

IV.  DISCUSSION

*A. Overriding the verbal and gestural responses*

A recurring feature in our corpus, observable in fragment 1, is that the content displayed in the ASR transcript was often responded to or used for a public performance, while the speaking turn produced by the robot was not dealt with. In sum, in case of trouble, the ASR transcript was often connected with a treatment of other modalities of expression of the robot (gestures, speech, etc.) as less relevant for the task at hand. In fragment 1, the robot's gestural or hearable conduct is not responded to and is overlapped by participants' reaction to the ASR transcript. Such conducts manifest an orientation to the ASR transcript as a sufficient contribution from the robot for participants to produce a response – response that the verbal and gestural behavior from the robot will not reconfigure a posteriori. The production of a verbal answer from the robot does not interrupt the reactions which follow the appearance of the ASR transcript, and, through this, does not reconfigure the manner in which participants demonstrably orient to the situation. This is unlike the symmetrically opposed situation, visible in e.g. fragment 2, where new information about the ASR transcript reconfigured the participant's interpretation of the previous verbal conduct from the robot. These observations suggest that the ASR transcript gives participants access to information which are more relevant for the practical problems (the "purposes at hand" [10]) they encounter in their situated interactions with the robot.

*B. Impact on sequential plasticity*

An additional dimension of the impact of the ASR transcript relates to the interactional competence of the robot. In HRI, many behaviors from a robot treated by co-presents members as a successful second pair part are, from the point of view of the engineers who programmed this robot, the result of an unforeseen sequence of events in which the robot [11], [12] did not, in fact, adapt to the human. Yet, when considering this regularly "fortuitous" character of the robot' successful responses [13] to the humans' actions, the ASR transcript appeared to often limit the "sequential plasticity" (Relieu et al.) of the robots' conduct by clearly indicating – for the numerous participants who treated it that way – what sequence of words the robot was responding to. In other terms, with a speech recognition transcript, there was a narrower range of behaviors from the robot which could be directly treated as social actions – that is, as responsive to the situated interaction and as "making relevant a set of potential next actions" [14].

We argue that, in the presence of an ASR transcript these behaviors from the robot become less simple to treat interactionally as stemming from an agent [15] imbued with intentionality. That is, more (incongruent) parameters have to be taken in account during the interactional work accomplished by some participants to preserve the appearance of "an intentional self" [16] to the robot. ASR transcripts limit the range of meaningful intentional patterns which can be connected with the robot's observable behavior. Such a context offers less resources to human participants to "safeguards the robot's status as an agent" [15] – for example, it may becomes more intricate for humans to attribute a mental state to the robot to account for a "missing action" on its part [15]. In our experiment, instead of hiding the absence of a cognitive "intersubjectivity" (or at least the absence of a correct receipt of human participants' speech) behind behaviors able to "mold in different sequential trajectories" [7], the Pepper robot provided additional resources to make these breakdowns ostentatious and interactionnally relevant.

The previous observations can be summarized with the following design dilemma: even though, as intended by the designers of Pepper, the ASR transcript did facilitate the identification and the repair of misunderstandings, it simultaneously limited the plasticity of the robot's behaviour and, through this, the smoothness of the interactions.

*C. The relevance of the ASR transcript for the robot itself*

The impact of the ASR transcript on participants' conduct highlights that, from the point of view of the robot itself, considering the ASR transcript is crucial to make sense of the interactions in which it is involved. In fragments analyzed in this work, human participants' reactions are only fully understandable if the presence of an ASR transcript in front of them is taken into account.

As [13] state, "a robot should not produce actions that make relevant next actions it will not be able to respond to", which, among other prerequisites, implies that the robot only "produce[s] actions which it is itself capable of understanding" [13]. In the situations we studied and with the conversational software we used, the ASR transcript contravenes [13]'s guideline: the display of a transcription of what the human said was an action from the robot that the robot was unable to make sense of. For example, in fragment 1, it was not possible to understand P1's utterance "mel" without taking in account that this word was written on the robot's tablet. If a robot were to attempt to interpret a participant's behavior without this crucial contextual element of "what it displayed on its tablet", it would often miss one of the most consequential features of these situated interactions. In other terms, it was critical for the robot to be able to identify when the conduct of human participants (repairs, accounts, repetitions, etc.) was not produced in reference to something it said (i.e. it is not an answer to a previous speaking turn) but, rather, in reference to something it was currently displaying (or had recently displayed).

These observations add to the list of key contextual parameters that embodied agents should perceive [17] and consider to better understand what human participants are indexing in their speaking turns. Yet, the relevance of the ASR transcript to participants also relates to the design question as to what robots should display of what they grasp from the world. Should the robot function as a complete blackbox, or should it display some of the information it uses to generate its actions? For example, should the robot display how many humans it currently sees, the confidence score attributed to the presence of each human, or even the confidence score attributed to each of the ASR results it shows? This question is already relevant for automatic cars [18], which usually provide for the passengers a feedback (among other possible ways to represent it) of what they "see" of the outside word: cars, pedestrians, motorcycles, lanes, etc.

## V. CONCLUSION: A COGNITIVIST DEFINITION OF MUTUAL UNDERSTANDING?

Based on the previous discussion, the orientation of many of our participants towards the ASR transcript can be described in two ways. A safe manner of verbalizing our participants' practices is to say that it became a members' problem that the ASR transcript displayed an adequate receipt of previous turns. The performance that was regularly expected from the robot was not only a gestural and a verbal one. It was also an "auditive" one. During their interactions, it generally mattered to participants when the robot displayed a too-dissimilar transcript of what they said. Yet, in the EMCA endeavor to represent as precisely as possible participants' own emerging categories during their situated activity, another level of description might be more faithful to the orientation displayed by many of our participants towards the ASR transcript: they were led to enact a cognitivist definition of mutual understanding. In other words, participants cared about what was inside the robot's "head" or "algorithm".

More precisely, as a technological artefact, we argue that the ASR transcript reified a specific representation of what human-human understanding is (in a conversation) which stands at the other end of the spectrum compared to the way it is conceptualized in ethnomethodology and conversation analysis. If, as Suchman mentions, "[e]very human tool relies upon, and reifies in material form, some underlying conception of the activity that it is designed to support." [19], the speechbar reified a conception of "understanding" as sharing a "similar mental representation about the world" [20] – rather than, as it is generally described in conversation analytic works, as "related to the next action achieved by the co-participant" ([6]; see also [21] for interspecies interactions).

Using a concept introduced by [22], we argue that the "script" inscribed in the ASR transcript – which stood out in front of participants each time they stopped speaking – encouraged an orientation to "what was inside the robot's head" as interactionally relevant: it facilitated a focus on the "hidden mental states and processes' (thought to be 'in our minds'), supposedly responsible for us doing what we do" [23], [24]. For those participants, it favored an orientation to progressivity in conversation as based on a perfectly shared mental reality, rather than as "being able to ''go on'' with each other" [24] or "to progress with whatever [participants] are doing together" [20].

[25] notes that "Medical imaging technologies, such as magnetic resonance imaging and ultrasound [...] make visible parts of the human body, or of a living fetus in the womb, that cannot be seen without them. But the specific way in which these technologies represent what they "see" helps to shape how the body or a fetus is perceived and interpreted and what decisions are made". Similarly, in the previous fragments, the ASR transcript made visible a specific aspect of the robot's perception of the world – however, unlike the ultrasound, without hiding the other dimensions from view. Through participants' verbal and embodied behavior (pointing the tablet, reading out loud what was written, etc.), this display from the robot of its perception of the word became a public and a consequential phenomenon. In the interpretation which we tried to outline here, these participants oriented towards the existence of a form of "shared reality" (at least about the phonological identification of what they said) as a requirement to progress with the interaction.

## APPENDIX

Transcriptions of talk follow [26]'s transcription conventions:

| | |
|---|---|
| = | Latching of utterances |
| (.) | Short pause in speech (<200 ms) |
| (0.6) | Timed pause to tenths of a second |
| : | Lengthening of the previous sound |
| . | Stopping fall in tone |
| ? | Rising intonation |
| °uh° | Softer sound than the surrounding talk |
| .h | Aspiration |
| h | Out breath |
| heh | Laughter |
| ((text)) | Described phenomena |
| (text) | Uncertain word |

Embodied actions were transcribed using [8]'s multimodal transcription conventions:

\* \* Gestures and descriptions of embodied actions are delimited between

+ + two identical symbols (one symbol per participant)

Δ Δ and are synchronized with corresponding stretches of talk.

\*--> The action described continues across subsequent lines

-->\* until the same symbol is reached.

\>> The action described begins before the excerpt's beginning.

\>> The action described continues after the excerpt's end.

p1 Participant doing the embodied action is identified in small caps in the margin

Symbols and abbreviations used in transcriptions referred to the following multimodal dimensions:

P1 Turn at talk from the human
RO Turn at talk from the robot
p1 Multimodal action from the human
ro Multimodal action from the robot
fi Screenshot of a transcribed event
£ Human's gaze
\* Human's arms
@ Human's whole body
^ Movement in space
$ Robot's arm
+ Robot's gaze
% Robot's belly screen (tablet)
# Position of a screenshot in the turn at talk

REFERENCES

[1] E. D. Kristiansen and G. Rasmussen, "Eye-tracking Recordings as Data in EMCA Studies: Exploring Possibilities and Limitations," *Social Interaction: Video-Based Studies of Human Sociality*, vol. 4, 2021, [Online]. Available: https://tidsskrift.dk/socialinteraction/article/view/121776
[2] A. Deppermann, "Inferential Practices in Social Interaction: A Conversation-Analytic Account," *Open Linguistics*, vol. 4, no. 1, pp. 35–55, Jan. 2018, doi: 10.1515/opli-2018-0003.
[3] A. T. Kerrison, "We're All Behind You: The Co-Construction of Turns and Sequences-at-Cheering," Ulster University, 2018.
[4] D. vom Lehn, "From Garfinkel's 'Experiments in Miniature' to the Ethnomethodological Analysis of Interaction," *Hum Stud*, vol. 42, no. 2, pp. 305–326, Sep. 2019, doi: 10.1007/s10746-019-09496-5.
[5] A. Deppermann, "How does 'cognition' matter to the analysis of talk-in-interaction?," *Language Sciences*, vol. 34, no. 6, pp. 746–767, 2012, doi: https://doi.org/10.1016/j.langsci.2012.04.013.
[6] L. Mondada, "Understanding as an embodied, situated and sequential achievement in interaction," *Journal of Pragmatics*, vol. 43, no. 2, pp. 542–552, 2011, doi: https://doi.org/10.1016/j.pragma.2010.08.019.
[7] M. Relieu, M. Sahin, and A. Francillon, "Une approche configurationnelle des leurres conversationnels," *Réseaux*, vol. N°220-221, no. 2, p. 81, 2020, doi: 10.3917/res.220.0081.
[8] L. Mondada, "Challenges of multimodality: Language and the body in social interaction," *Journal of Sociolinguistics*, vol. 20, no. 3, pp. 336–366, Jun. 2016, doi: 10.1111/josl.1_12177.
[9] E. A. Schegloff, "Repair After Next Turn: The Last Structurally Provided Defense of Intersubjectivity in Conversation," *American Journal of Sociology*, vol. 97, pp. 1295–1345, 1992.
[10] H. Garfinkel, *Studies in Ethnomethodology*. Cambridge: Polity Press, 1967.
[11] H. R. M. Pelikan, M. Broth, and L. Keevallik, "'Are You Sad, Cozmo?': How Humans Make Sense of a Home Robot's Emotion Displays," in *Proceedings of the 2020 ACM/IEEE International Conference on Human-Robot Interaction*, New York, NY, USA: Association for Computing Machinery, 2020, pp. 461–470. [Online]. Available: https://doi.org/10.1145/3319502.3374814
[12] D. Rudaz, K. Tatarian, R. Stower, and C. Licoppe, "From Inanimate Object to Agent: Impact of Pre-Beginnings on the Emergence of Greetings with a Robot," *J. Hum.-Robot Interact.*, vol. 12, no. 3, Apr. 2023, doi: 10.1145/3575806.
[13] S. Tuncer, C. Licoppe, P. Luff, and C. Heath, "Recipient design in human–robot interaction: the emergent assessment of a robot's competence," *AI & Soc*, Jan. 2023, doi: 10.1007/s00146-022-01608-7.
[14] S. Tuncer, S. Gillet, and I. Leite, "Robot-Mediated Inclusive Processes in Groups of Children: From Gaze Aversion to Mutual Smiling Gaze," *Frontiers in Robotics and AI*, vol. 9, 2022, doi: 10.3389/frobt.2022.729146.
[15] H. Pelikan, M. Broth, and L. Keevallik, "When a Robot Comes to Life: The Interactional Achievement of Agency as a Transient Phenomenon," *SI*, vol. 5, no. 3, Oct. 2022, doi: 10.7146/si.v5i3.129915.
[16] M. Alač, Y. Gluzman, T. Aflatoun, A. Bari, B. Jing, and G. Mozqueda, "Talking to a Toaster: How Everyday Interactions with Digital Voice Assistants Resist a Return to the Individual," *Evental Aesthetics*, vol. 9, 2020, [Online]. Available: https://eventalaaesthetics.net/wp-content/uploads/2021/03/EAV9N1_2020_Alac_Toaster_3_53.pdf
[17] A. Vinciarelli, M. Pantic, and H. Bourlard, "Social signal processing: Survey of an emerging domain," *Image and Vision Computing*, vol. 27, no. 12, pp. 1743–1759, Nov. 2009, doi: 10.1016/j.imavis.2008.11.007.
[18] B. Brown, M. Broth, and E. Vinkhuyzen, "The Halting problem: Video analysis of self-driving cars in traffic," in *Proceedings of the 2023 CHI Conference on Human Factors in Computing Systems*, Hamburg Germany: ACM, Apr. 2023, pp. 1–14. doi: 10.1145/3544548.3581045.
[19] L. A. Suchman, *Plans and situated actions: The problem of human-machine communication*. in Plans and situated actions: The problem of human-machine communication. New York, NY, US: Cambridge University Press, 1987, pp. xii, 203.
[20] S. Albert and J. Ruiter, "Repair: The Interface Between Interaction and Cognition," *Topics in Cognitive Science*, vol. 10, pp. 279–313, Apr. 2018, doi: 10.1111/tops.12339.
[21] C. Mondémé, "Why study turn-taking sequences in interspecies interactions?," *J Theory Soc Behav*, vol. 52, no. 1, pp. 67–85, Mar. 2022, doi: 10.1111/jtsb.12295.
[22] M. Akrich, "La construction d'un système socio-technique. Esquisse pour une anthropologie des techniques," *Anthropologie et Sociétés*, vol. 13, Jan. 1989, doi: 10.7202/015076ar.
[23] J. Shotter, "'Now I can go on:' Wittgenstein and our embodied embeddedness in the 'Hurly-Burly' of life," *Hum Stud*, vol. 19, no. 4, pp. 385–407, Oct. 1996, doi: 10.1007/BF00188850.
[24] L. Sterponi and A. Fasulo, "'How to Go On': Intersubjectivity and Progressivity in the Communication of a Child with Autism," *Ethos*, vol. 38, pp. 116–142, Mar. 2010, doi: 10.1111/j.1548-1352.2009.01084.x.
[25] P.-P. Verbeek, "Materializing Morality: Design Ethics and Technological Mediation," *Science, Technology, & Human Values*,